# Linear dichroism of the optical properties of SnS and SnSe van der Waals crystals

Agata K. Tołoczko,*[a] Jakub Ziembicki,[a] Miłosz Grodzicki,[ab] Jarosław Serafińczuk,[ab] Seth A. Tongay,[c] Melike Erdi,[c] Natalia Olszowska,[d] Marcin Rosmus,[d] and Robert Kudrawiec[ab]

[a] Department of Semiconductor Materials Engineering, Wrocław University of Science and Technology, Wybrzeże Wyspiańskiego 27, 50-370 Wrocław, Poland
[b] Łukasiewicz Research Network – PORT Polish Center for Technology Development, Stabłowicka 147, Wrocław, Poland
[c] School for Engineering of Matter, Transport and Energy Arizona State University Tempe, AZ 85287, USA
[d] Solaris National Synchrotron Radiation Centre, Jagiellonian University, Czerwone Maki 98, 30-392 Kraków, Poland

Tin monochalcogendies SnS and SnSe, belonging to a familiy of van der Waals crystals isoelectronic to black phosphorus, are know as enivornmetally-friendly materials promisng for thermoelecric conversion applications. However, they exhibit other desired functionalities, such as intrisic linear dichroism of the optical and electronic properties originating from strongly anisotropic orthorhombic crystal structure. This property makes them perfect candidats for polarization-sensitive photodetectors working in near infrared spectral range. We present a comprehensive study of the SnS and SnSe crystals by means of optical spectroscopy and photoemission spectroscopy, supported by *ab initio* calcualtions. The studies revealed the high sensitivity of the optical response of both materials to the incident light polarization, which we interpret in terms of the electronic band dispersion and orbital composition of the electronic bands, dictating the selection rules. From the photoemission investigation we determine the ionization potential, electron affinity and work function, which are parameters crucial for the design of devices based on semiconductor heterostructures.

## Introduction

The unique layered structure and possibility of obtaining atomically thin flakes make van der Waals (vdW) crystals[1,2] perfect candidates for applications in two-dimensional electronics and optoelectronics, such as photodetectors, solar cells, and light emitters. An interesting class of devices are polarization-sensitive photodetectors, as their responsivity depends on the incident light polarization. Such technology can be exploited for detection of light polarization changes after traveling through a birefringent medium, including various solids, liquid crystals, but also biological systems, such as protein solutions.[3] In the case of the latter, the measured polarization angle shift may allow to determine presence of certain proteins in a sample, which is extremally important in diagnostics. Unfortunately, the state-of-the-art polarization-sensitive photodetectors are rather complex, as they demand integration of multiple optical components. The design, however, can be significantly simplified by exploiting the materials exhibiting intrinsic anisotropy of the optical properties.[4–6] Among vdW crystals, such properties was observed for black phosphorus (BP)[6–8] and its binary analogues, group IV monochalcogenides (MX, where M = Ge, Sn, and X = S, Se).[9–16] Since MXs characterize with stability in atmospheric conditions superior to BP,[9,17] they are more suitable for most applications, including photodetection. The origin of the anisotropy is the structure of the materials. Similarly to BP, MXs crystallize in a distorted orthorhombic structure (space group *Pnma*, no. 62) with strong in-plane anisotropy of the atomic arrangement, as schematically presented in Fig. 1a. Along the *x* axis the structure is puckered, while a ladder-like pattern is observed in the *y* direction. The anisotropy of the crystal lattice induces directionality of the electronic band dispersion, which then determines the dielectric function and the optical properties. For the considered 1:1 stoichiometry, the metal atom is in the +2 oxidation state and forms three bonds with chalcogen atoms and a lone electron pair in the tetragonal coordination.[18] The presence of the stereochemically active lone pairs is also know to reduce the ionization potential of a material[19–21] and influence the photoconversion efficiency, as in the case of metal-halide perovskites.[22,23]

Among the MX family, tin monochalcogenides characterize with relatively narrow indirect band gaps in the near infrared spectral range (1.1 and 0.9 eV for SnS and SnSe, respectively),[24,25] high absorption coefficient, and intrinsic *p*-type conductivity, enhanced by the presence of native acceptor defects,[26–30] making them perfect candidates for infrared polarization-sensitive photodetectors. Apart from optoelectronics, SnX crystals poses great potential for applications as thermoelectric materials,[31–34] extremally desired in the era of growing energy demand. For SnSe, Zhao *et al.*[31] reported unexpectedly high thermoelectric figure of merit of 2.6, exceeding the performance of typical state-of-the-art Pb-based materials.[35]

In this work we investigate the linear dichroism of the optical properties of SnS and SnSe by means of optical spectroscopy and search for its origins in the electronic band structure, studied by combined density functional theory (DFT) calculations and photoemission spectroscopy. We discuss the influence of the stereochemical activity of the lone electron pairs on the electronic band energies with respect to the vacuum level, crucial for the engineering of semiconductor heterostructures and metal contacts.[27,36–39] The obtained results provide in-depth understanding of the mechanisms responsible for the observed phenomena at the fundamental level and reflect the empirical implications of the selection rules.

## Results and discussion

### a. Structural characterization

SnS and SnSe crystallize in a distorted orthorhombic phase, schematically presented in Fig. 1a. In Fig. 1b the first Brillouin zone (BZ) of the reciprocal space is illustrated, with marked high-symmetry points. The structure of the investigated samples was confirmed by X-ray diffraction measurements (XRD), revealing characteristic reflexes, assigned according to the PDF-4 database[40] (SnS card no. 00-067-0519, SnSe card no. 00-048-1224), as shown in Fig. 1c. For both crystals, a dominant [001] orientation was observed in the diffraction patterns. In addition, a reflection from Si (400) plane was detected. The lattice parameter $c$ (corresponding to the out-of-plane crystallographic direction) equals 11.18 and 11.48 Å for SnS and SnSe, respectively, in agreement with previous reports.[41]

An investigation of the chemical composition was performed using core level X-ray photoemission spectroscopy (XPS) (Fig. 1d,e). The full-scale spectra, plotted in Fig. S1 of the Electronic Supplementary Information (ESI), only show the presence of Ge or Sn and S or Se atoms, with negligible traces of oxygen and carbon, indicating high purity of the sample surface. In Fig. 1d,e regions of the spectra with characteristic features related to contributing elements are presented, for SnS (panel d) and SnSe (panel e), identified after Moulder et al.[42] The energies of individual orbital levels were determined by fitting the Gaussian line-shape to the acquired data. For S in SnS a doublet corresponding to $2p$ orbitals was observed, with the expected $p_{3/2} : p_{1/2}$ intensity ratio of ~2:1. For Se in SnSe, a $3d$ doublet can be identified, with the $d_{5/2} : d_{3/2}$ intensity ratio of ~3:2. For Sn atoms, signal associated with $3d$ orbitals was detected. The exact energies and spin-orbit splitting of the observed lines are summarized in Table S1 of the ESI. Along with the core levels corresponding to the SnX phase (with Sn at the +2 oxidation state) plotted as red dashed components of the Gaussian fits, a weak signal originating from the SnX$_2$ phase (Sn at the +4 oxidation state, yellow dashed lines) was observed as an asymmetrical broadening of the XPS lines. The contribution of the side-lines could not be resolved for the SnSe Se $3d$ peak due to close position of the $d_{3/2}$ and $d_{5/2}$ components, but is apparent for other measured lines, indicating presence of chalcogen-rich domains at the sample surface, which is not unusual (tin vacancies are one of the most stable native defects in the system, responsible for the intrinsic p-type character).[43–45]

### b. Optical properties

The optical properties of SnS and SnSe crystals were investigated by means of complementary methods of optical spectroscopy: photoreflectance (PR), sensitive to direct optical transitions,[46–48] and optical absorption allowing to detect both direct and indirect band gap. In Fig. 2 a comparison of the photoreflectance and optical absorption spectra acquired at the temperature of 20 K for SnS (Fig. 2a) and SnSe (Fig. 2c) is presented. For both materials, in the PR spectra (plotted with blue and purple circles) multiple resonances are visible, indicating contribution of three and four optical transitions for SnS and SnSe, respectively, labelled $E_1$-$E_4$ in the figure. In order to determine the transitions energies, the Aspnes formula,[49] given by

$$\frac{\Delta R}{R}(\hbar\omega) = Re\left(\sum C_i e^{i\varphi_i}(\hbar\omega - E_i + i\Gamma_i)^{-2}\right) \quad (1)$$

was fitted to the experimental data. In the equation, $C_i$ is the amplitude of the $i$-th PR resonance, $\varphi_i$ is the phase, $\Gamma_i$ is the broadening and $E_i$ is the energy. The energies corresponding to the temperature of 20 K are summarized in Table 1. Based on the fitting parameters, the resonance moduli, plotted as shaded areas in Fig. 2a,c, were calculated, using the formula

$$\Delta\rho_i(E) = \frac{|C_i|}{(E-E_i)^2 + \Gamma_i^2}. \qquad (2)$$

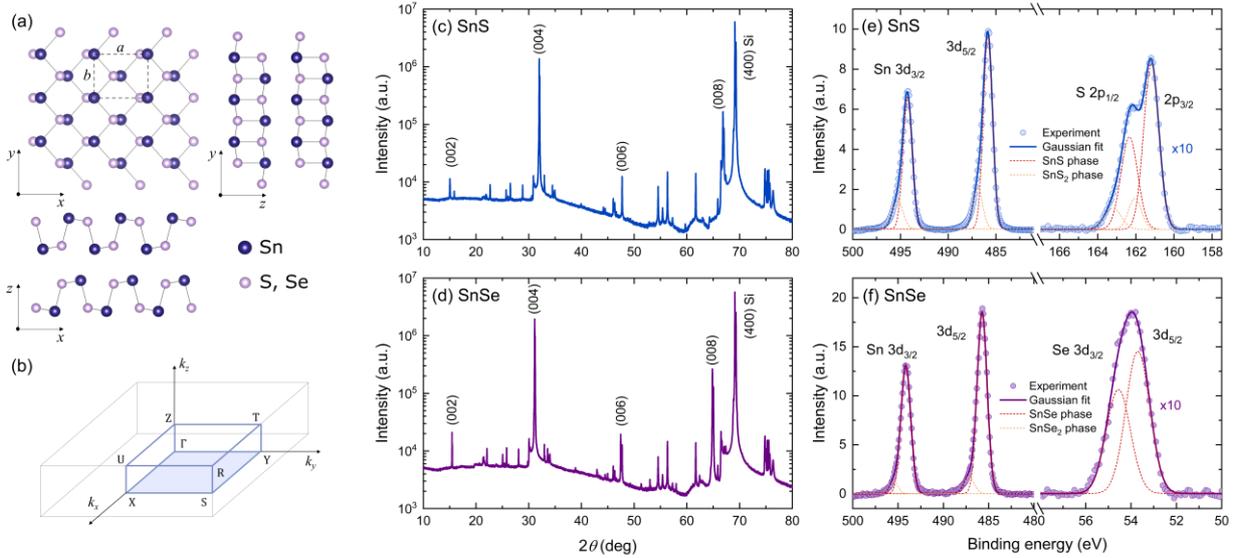

**Fig. 1.** Schematic illustration of the crystal structure (a), the first Brillouin zone of the reciprocal orthorhombic structure (b), the results of XRD characterization for SnS (c) and SnSe (d), and core level XPS spectra of SnS (e) and SnSe (f). In panels e and f the Gaussian line-shape (dark solid lines) were fitted to the experimental points (circles). The components of the Gaussian fits corresponding to SnX and SnX$_2$ phases are plotted with dashed red and yellow lines, respectively.

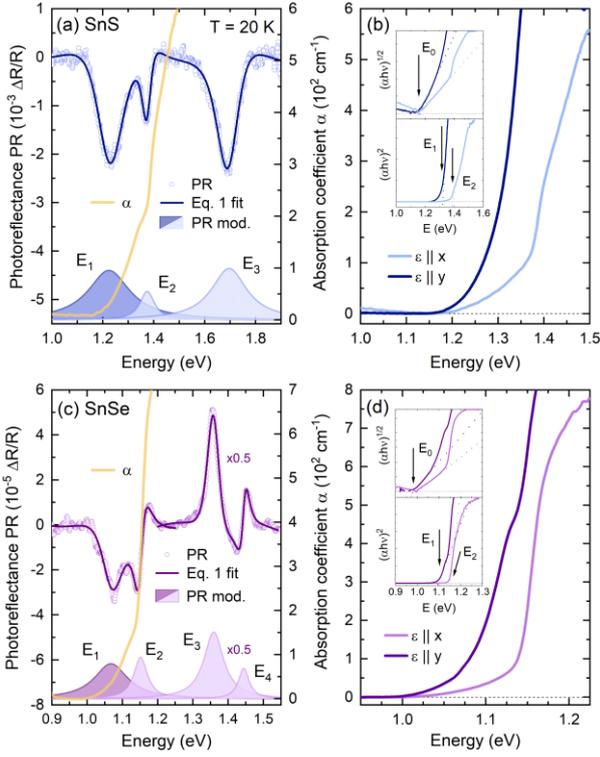

**Fig. 2.** The unpolarized photoreflectance and optical absorption spectra acquired for SnS (a) and SnSe (c). The fit with the Aspnes formula (dark solid lines) is superimposed over the experiemental PR data (circles). The absorption coefficient α is plotted with solid yellow lines. Shaded areas at the bottom of each panel are the PR resonances moduli (Eq. 2), with the dark and light color corresponding to $y$ and $x$ polarization of the optical transition, respectively. In panel b, the $E_3$ and $E_4$ rezonances and corresponding moduli are scaled by the factor of 0.5 for clarity. The optical absorption spectra measured at the incident light polarized along $x$ (bright solid lines) and $y$ (dark solid lines) crystallographic direction for SnS (b) and SnSe (d). In the insets Tauc plots for indirect and direct absorption edge are shown.

The area under the modulus curve is related to the transition oscillator strength.

The PR resonances visible in the spectra exhibit strong polarization dependence, as shown by Ho *et al.*[50] and Herninda *et al.*[51] and confirmed by our preliminary results. The $E_1$ transition for both SnS and SnSe is polarized along the $y$ direction, while all the energetically higher features manifest the $x$ polarization, as illustrated by different shades of the moduli (dark for $y$ and light for $x$ polarization) of the measured PR resonances.

The optical absorption coefficient spectra (yellow solid lines in Fig. 2a,c) exhibit a characteristic step-like shape, also suggesting contribution of multiple optical transitions, with the two branches corresponding well to the $E_1$ and $E_2$ PR resonances. Above ~1.5 eV for SnS and ~1.3 eV for SnSe the measured signal is saturated and no features attributed to the higher transitions could be observed, however below the first resonance energy, an absorption tail is present, which may be an evidence of the indirect character of the fundamental band gap. The indirect absorption edge could be better resolved in the spectra acquired for the incident light polarized along one of the main crystallographic directions (i.e. the electric component of the EM field $\varepsilon \parallel x$ or $\varepsilon \parallel y$), presented in Fig. 2b,d. A significant shift of the direct absorption edge with varying polarization can be observed, related to the changes of the probability of individual transitions, in-line with the PR resonances polarization dependence. From the Tauc plots of the absorption coefficient spectra, presented in the insets of panels b and d, the energies of the individual optical transitions were extracted. For the $x$ polarization the linear region in the $(\alpha h\nu)^{1/2}$ plot is clearly visible, allowing to determine the fundamental indirect band gaps $E_0$ of 1.16 eV for SnS and 0.99 eV for SnSe. The $(\alpha h\nu)^2$ plots of the energetically higher regions of the spectra for both polarizations provided the energies of direct transitions $E_1$ and $E_2$, in perfect agreement with PR, as compared in Table 1. Some minor discrepancies may result from the fact that the Tauc plot method is best applicable for materials with simple electronic band dispersion, exhibiting the absorption edge originating from an optical transitions between well-defined parabolic-like valleys.

To better understand the optical activity of SnXs, temperature dependent experiments were carried. The obtained results are presented in Fig. 3. The temperature evolution of the PR spectra (Fig. 3a,d) reveals the expected thermal redshift of the PR resonances energies and a decrease of their amplitude. In the case of optical absorption, the influence of the temperature was investigated using either unpolarized light (Fig. 3b,e) or light polarized linearly along $x$ and $y$ direction (Fig. S2 of the ESI).

Temperature dependences of the $E_0$ and $E_1$ transitions energies (Fig. 3c,f) were approximated by Bose-Einstein (B-E)[52] and Varshni[53] formulas, providing information about the electron-phonon interaction strength. The equations and obtained values of the fitting parameters values are given in the ESI (Table S2). Both approaches allow to reproduce the temperature dependence with good accuracy, although the B-E procedure is more accurate at low temperatures. For the energetically higher transitions the fits did not converge or provided non-physical values, which may be related to the uncommon shape of the plot, resulting from the uncertainties of the energies extracted from the PR spectra, especially at higher temperatures.

**c. Electronic band structure**

Based on the absolute energies and polarizations of the optical transitions observed in the experiment, we propose an assignation to certain Brillouin zone points. In Fig. 4a,c the electronic band structure is presented, calculated employing DFT, with the use of Heyd-Scuseria-Ernzerhof (HSE06) hybrid functional (for the computational details see *Methods* section). The calculations confirm the multivalley character of the band dispersion, with the conduction band minimum (CBM) in the Γ-Y path and a characteristic shape of the valence band maximum (VBM) in the Γ-X path, composed of two hole pockets. The band shape has been observed experimentally in the angle-resolved photoemission spectroscopy measurements and strongly influences the thermoelectric efficiency of the materials. The band dispersion is generally similar to the germanium monochalcogenides (GeS, GeSe), except for the valence band in the Γ point, pushed towards higher binding energies relative to the VBM.[13,14,54,55] Hence, the quasi-direct band gap character observed for GeS[54] and GeSe[13,14] is not present for SnS and SnSe crystals. In Fig. 4b,e the relative distance between highest valence band and lowest conduction band is plotted, along with in-plane components of the transition squared matrix element (shaded areas), determining the transition oscillator strength and, as a consequence, its probability. The matrix element distribution is governed by the selection rules and indicates for which polarization a transition is allowed. By comparing the experimental results to the calculated band structure we assign the observed optical transitions as labelled in Fig. 4b,e and summarized in Table 1. The $E_0$ transition can be attributed to the indirect fundamental band gap between VBM and CBM. The lowest direct transition $E_1$, polarized along $y$ direction, correspond to the valley in the Γ-Y path, while $E_2$, active for the $x$ polarization, occurs in the minimum in the Γ-X path. The transition $E_3$ can be assigned to the U point, and $E_4$, visible only for SnSe, to a critical point in the Y-S path. The agreement between experimental and theoretical energies is excellent for the $E_0$, but worse for the higher transitions. Such discrepancies are not unusual, as DFT calculations, burdened with approximations and sensitive to the computational parameters, often reproduce the band shapes accurately, but fail to evaluate the absolute energies. Therefore, to find a plausible interpretation of the experimental results we focus mainly on the polarization and the matrix element in-plane components ratio.

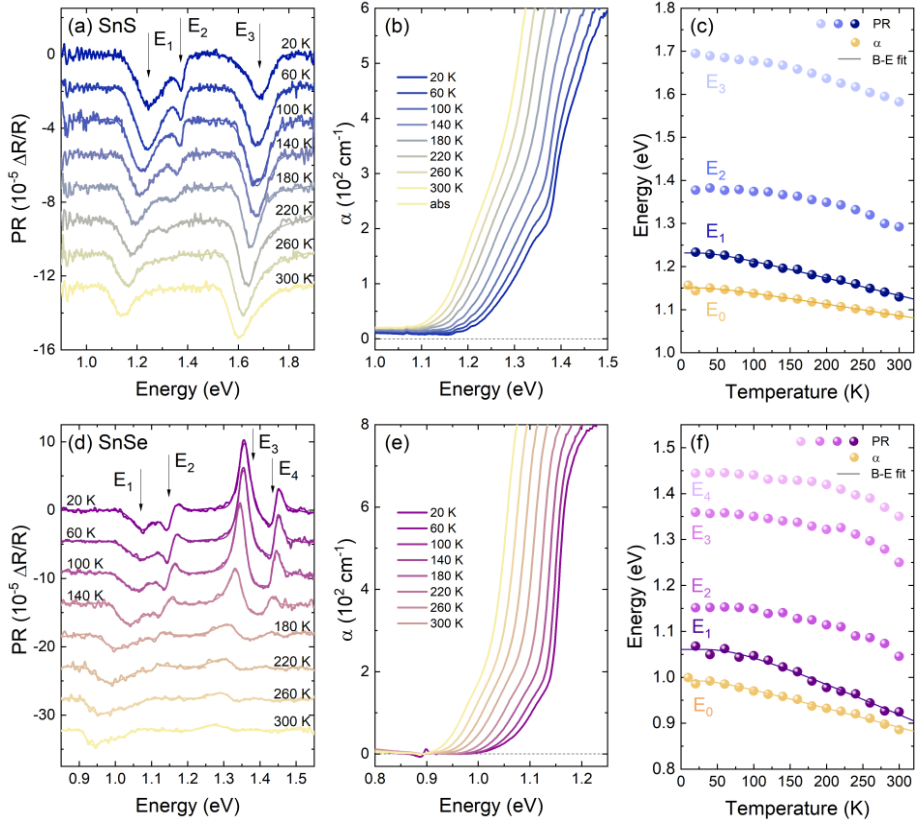

**Fig. 3.** The temperature evolution of the photoreflectance (a,d) and optical absorption (b,e) spectra acquired for SnS (a,b) and SnSe (d,e). Temperature dependence of the optical transition energies determined based on the optical absorption ($E_0$) and PR ($E_1$ - $E_4$) measurements for SnS (c) and SnSe (f). The experimental data (circles) are approximated by Bose-Einstein formula (solid lines).

The anisotropy of the matrix element distribution and, as a consequence, the polarization of the optical transitions is related to the orbital composition of the electronic bands. In Fig. 4c,f the density of states (DOS) is plotted, including both total DOS (shaded areas) and partial contribution of the Sn and X valence orbitals (dashed lines): Sn 5$s$, 5$p$ and S 3$s$, 3$p$ or Se 4$s$, 4$p$. According to DFT calculations, also $d$ orbitals (often referred to as semi-core states) have minor contribution to the total DOS, which is not plotted in the figure for clarity. It was shown that either including the $d$ states to the band structure calculation, or treating them as core-levels, provides nearly identical results.[19,20] The conduction band of both materials is mainly composed of Sn 5$p$ orbitals, with addition of X $p$ states. In the valence band three regions can be distinguished, labelled A, B, and C in Fig. 4c,f, as was also shown for other MXs.[20,51] The A band is composed of a mixture of Sn 5$s$, 5$p$ and X $p$ orbitals, the B region is dominated by X $p$ states, with contribution of Sn $p$ orbitals, and the C peak almost entirely consists of Sn 5$s$ states. The contribution of the 5$s$ orbitals to the upper valence band is a consequence of the presence of stereochemically active $s^2$ lone electron pairs in the crystal structure and will be discussed further in the text. In Fig. 4a,d, the contribution of the three spatial components of the $p$ orbital ($p_x$, $p_y$, $p_z$), oriented along respective crystallographic axes, is superimposed over the band dispersion. In the figure the $p$ states of both Sn and X atoms are combined. More detailed picture, including also the $s$ states and resolving contribution of different elements, is presented in Fig. S3 of the ESI. In the orbital composition plots the size of the points is proportional to the contribution of an orbital to certain band. The distribution of the three components across the BZ is particularly interesting regarding the topmost valence and lowermost conduction band, as it affects the optical properties observed in the experiment. In the proximity of the X and U points of the BZ, the bands are mainly composed of $p_x$ orbitals, resulting in high $k_x$ matrix element component, and consequently polarization of the optical transition along $x$ direction. Similarly, in the Γ-Y and Z-T paths the $p_y$ orbitals dominate, giving rise to high $k_y$ matrix element component and determining the $y$ polarization of the optical transitions. It can also be seen that in the BZ regions where both valence and conduction bands consists mainly of $p_z$ states (Γ, Z, S, and R points), the in-plane matrix element vanishes, as a result of small value of the overlap integral. Therefore, in the applied experimental configuration, no optical transition attributed to the Γ point of the BZ was observed, despite relatively low energy with respect to the fundamental band gap.

**Table 1.** Experimental and theoretical energies and polarizations of measured and predicted optical transitions, along with their assignation to certain BZ points. The experimental values correspond to the temperature of 20 K.

|  | Transition | BZ point | Polarization | Energy (eV) | | |
|---|---|---|---|---|---|---|
|  |  |  |  | Experiment (20 K) | | DFT (0 K) |
|  |  |  |  | Absorption | PR |  |
| SnS | $E_0$ | Γ-X→ Γ-Y | - | 1.16 | - | 1.16 |
|  | $E_1$ | Γ-Y | $y$ | 1.23 | 1.23 | 1.39 |
|  | $E_2$ | Γ-X | $x$ | 1.38 | 1.38 | 1.75 |
|  | $E_3$ | U | $x$ | - | 1.70 | 1.97 |
| SnSe | $E_0$ | Γ-X→ Γ-Y | - | 0.99 | - | 0.95 |
|  | $E_1$ | Γ-Y | $y$ | 1.09 | 1.07 | 1.16 |
|  | $E_2$ | Γ-X | $x$ | 1.14 | 1.15 | 1.31 |
|  | $E_3$ | U | $x$ | - | 1.36 | 1.63 |
|  | $E_4$ | Y-S | $x$ | - | 1.45 | 1.72 |

### d. Photoemission study

The valence band DOS can be experimentally investigated by means of photoemission spectroscopy. In this study we apply two techniques, exploiting different radiation sources and excitation energies. The lab-based XPS, utilizing the Al K$_\alpha$ line of $hv$ = 1486.6 eV, allows observation of core-level states, confirming the material composition and quality, and the high energy secondary electron cut-off, providing direct information about the work function of the investigated sample. In terms of valence band investigation, for high excitation energies the photoionization cross-section of the valence orbitals is relatively small, resulting in low photoemission intensity. Therefore, the second applied method was UV photoemission spectroscopy (UPS), exploiting monochromatic synchrotron radiation of $hv$ = 100 eV. The technique and the experimental setup provide significantly better resolution and sensitivity, along with possibility of angle-resolved measurements, which are reported in our previous work.[55] Both techniques allow measurements of the valence band, however UPS is more suitable, considering the photoionization cross-section of the valence orbitals, up to two orders of magnitude higher for the excitation energy of 100 eV compared to 1.5 keV.[56,57] In Fig. 5a,c the UPS spectra (top plot of each panel) are compared with the simulated valence band (bottom plots), obtained by applying following procedure to the calculated partial DOS: first, the contribution from the orbitals was weighted using the photoemission cross-sections corresponding to the excitation energy of 100 eV, according to Yeh and Lindau[56]. Next, to introduce the broadening of the electronic states, the partial DOS was convolved with Lorenzian (natural broadening) and Gaussian (thermal broadening) functions. The UPS spectra were corrected by subtracting the background originating from the inelastic electron scattering. It can be seen for both SnS and SnSe that after taking into account the photoionization cross-section, the signal from A and B bands (as labelled in Fig. 4c,d) is strongly dominated by X $p$ orbitals, and the contribution of Sn 5$s$ states is distinct only in the C band. The simulated curves are in perfect agreement with the experimental spectra in the B and C regions, but diverges in the A region. In the calculations, the intensity of the A band is lower with respect to the B band, but in the experiment we observe opposite relation. The discrepancy may result from the fact that in the measurement light polarized linearly was used, interacting differently with the three spatial components of the $p$ orbitals. Effectively, the photoionization cross-section for $p$ states might be smaller than predicted, while estimated correctly for $s$ orbitals, insensitive to the polarization. Then, the contribution of the Sn 5$s$ states to the photoemission from the A region should be greater, resulting in higher overall detected intensity. Another factor which may influence the respective bands intensity ratio is the possibility that the DFT calculations underestimate the contribution of the Sn 5s states to the topmost region of the valence band.

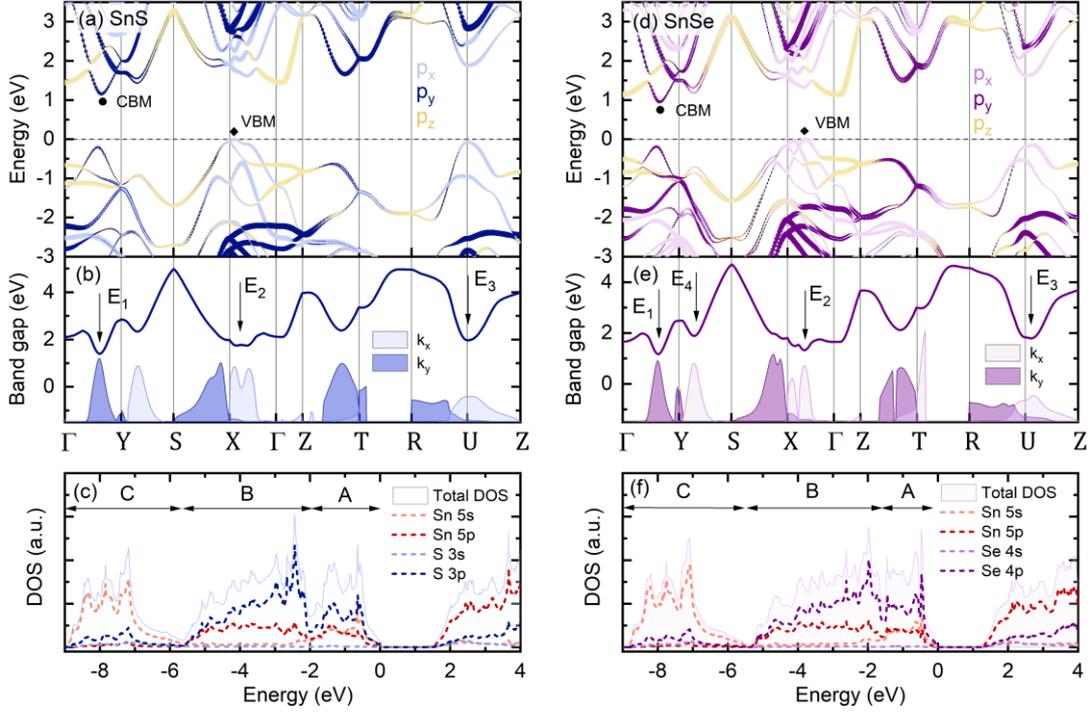

**Fig. 4.** (a, d) The electronic band dispersion of SnS (a) and SnSe (d) along high-symmetry BZ paths with superimposed contribution of the x, y, and z components of the Sn and X p orbitals. (b,e) The interband distance (solid line) and transition squared matrix element $k_x$ and $k_y$ components (bright and dark shaded areas, respectively). The optical transitions observed in the experiment are assigned to certain BZ points by arrows. (c,f) The calculated total (shaded areas) and partial (dashed lines) density of states, corresponding to individual orbitals. The energy scale is relative to the VBM energy ($E_{VBM}$ = 0). In panels a, b, and c results for SnS are presented, while d, e, and f correspond to SnSe.

The XPS spectra acquired for SnS and SnSe, visualizing the secondary electron cut-off and valence band maximum are presented in Fig. 5b,d. From the extrapolation of the linear region at high binding energy side of the spectrum the cut-off energy $E_{cut\text{-}off}$ was determined. The valence band maximum energy $E_{VBM}$ can also be estimated (with respect to the Fermi level), however considering the spectral resolution and intensity of the signal originating from the valence band, more reliable and precise values of $E_{VBM}$ are extracted from the UPS data, as shown in the insets of Fig. 5a,c.

It can be noticed that the VBM is observed at slightly higher energies in the XPS measurements. Such shift can be explained by the fact that the DOS at VBM (close to the X point of the BZ) is low and cannot be resolved in the XPS measurements. In the UPS spectra a characteristic step attributed to the VBM is clearly visible, followed by a rapid rise of the photoemission signal originating from a band in the Γ point.
The values of $E_{VBM}$ obtained for SnS and SnSe are 0.33 eV and 0.12 eV, respectively, remaining in good agreement with character of the crystals. The material work function ($\varphi$) can be calculated from the relation

$$\varphi = h\nu - E_{cut-off}. \qquad (3)$$

Then, the ionization potential (IP) is given by

$$IP = \varphi + E_{VBM}, \qquad (4)$$

and the electron affinity ($\chi$) can be defined as

$$\chi = IP - E_g, \qquad (5)$$

where $E_g$ is the fundamental energy gap determined from the optical absorption measurements. The values of the above parameters for SnS and SnSe (compared with GeS and GeSe investigated in our previous work[59]) are summarized in Table 2 and Fig. 6, visualizing the band alignment of the four MXs with respect to the vacuum energy. It should be noted that the work function, related to the Fermi level position inside the band gap, may vary depending on the intrinsic defect or intentional dopants concentration, therefore the parameter more suitable for comparison with other studies is the IP, i.e. the VBM position with respect to vacuum level.

The obtained values of the IP are relatively low compared to other semiconducting van der Waals crystals, which leads to some significant consequences considering potential applications, such as band alignment with window layers for photovoltaics, affecting the device efficiency, or with metals for electrical contacts, determining the ohmic or Schottky character of the junction. The reason for this is the aforementioned contribution of the metal $s^2$ lone electron pairs to the high energy region of the valence band, confirmed by previous studies.[19,60] In all MXs the metal atom is in the +2 oxidation state. For group IV metals it implies either tetrahedral or octahedral coordination. The former characterize with three bonds with chalcogen atoms and one lone electron pair in the fourth vertex of the tetrahedron, forming a distorted orthorhombic crystal lattice. In the latter, six M-X bonds result in perfectly symmetric rocksalt structure.[18,21] The lattice distortion is related to the stereochemical activity of the $s^2$ lone pairs, observed for GeS, GeSe, GeTe, SnS and SnSe, but not for other group IV-VI compounds with the same stoichiometry, such as SnTe, PbS, PbSe, and PbTe. The phenomenon can be partly explained by the revised lone pair model proposed by Walsh *et al.*[21] In the considered materials the metal valence $s$ electrons hybridize with chalcogen $p$ states, forming bonding and anti-bonding states (as schematically illustrated in Fig. 7), contributing to the high (C band) and low (A band) binding energy regions of the valence band, respectively. The amount of the contribution of the M $s$ orbitals to the anti-bonding state is determined by the relative energy of the M $s$ and X $p$ levels, yielding higher contribution for smaller energy distance. For the strong M $s$ component, the interaction of the unoccupied M $p$ orbitals with the anti-bonding state leads to asymmetric electron density distribution with a directional lone electron pair. When M $s$ contribution is minor, the interaction is weak and stereochemically active lone pairs do not form. The relative positions of the M $s$ and X $p$ valence levels for group IV and VI elements is presented in a diagram in Fig. 7, with the energies adapted after Mann *et al.*[61] It can be seen that due to lowest energy distance, most prominent lone pairs form for oxides, and the X $p$ energy increases down the VI group. For metal atoms, the Sn 5$s$ state is positioned higher relative to both Ge 4$s$ and Pb 6$s$ levels, reducing the effective distance to the X $p$ states, which explains the significantly lower IP observed for SnX compared to GeX. Considering only tin compounds, we should expect stronger mixing of the Sn 5$s$ orbitals with S 3$p$ than with Se 4$p$ states, and therefore greater contribution to the A band for SnS. In Fig. S5 of the ESI the Sn 5$s$ contribution to the total DOS is plotted for both investigated materials, confirming the predictions.

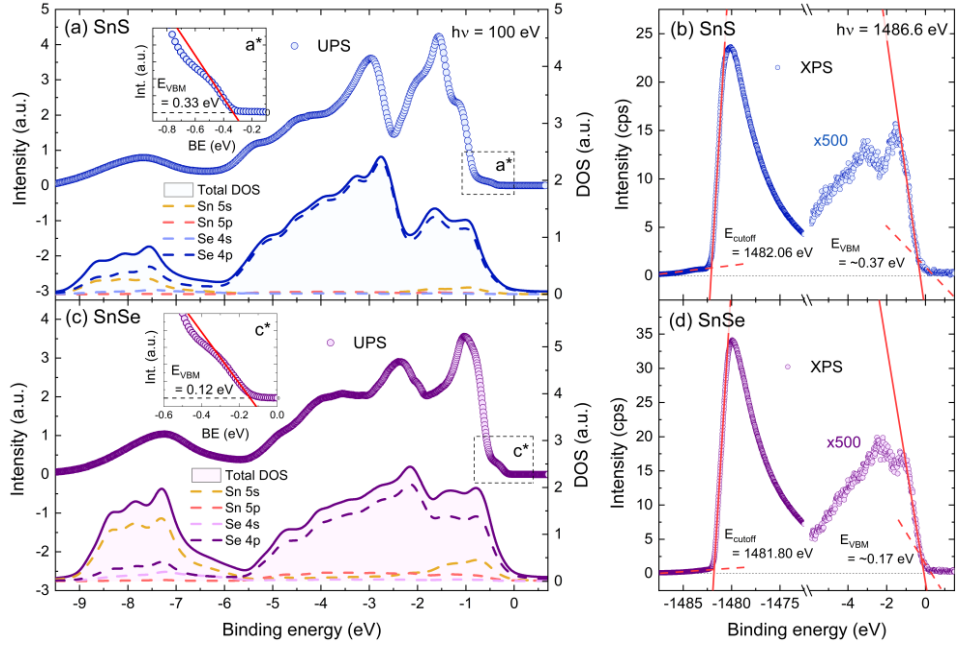

**Fig. 5.** (a,c) The results of the UPS measurements (circles, top of each panel) acquired with the use of synchrotron radiation at the excitation energy of 100 eV, compared with the simulated valence band (solid and dashed lines, bottom of each panel) based on the calculated DOS, corrected by the photoionization cross-section for individual orbitals and broadening of the electronic states. In the insets the boxed regions of the experimental spectra labeled a* and c* are enlarged to better illustrate the VBM and determine $E_{VBM}$. (b,d) The low (valence band) and high (secondary electron cut-off) binding energy regions of the lab-based XPS spectra (acquired with the use of Al K$_\alpha$ line of 1486.6 eV), allowing to extract the $E_{cut-off}$ and $E_{VBM}$ (less accurate than from UPS spectra). The valence band regions of the spectra are scaled by the factor of 500 for better visualization.

The revised lone pair model is in agreement with the experimental results obtained for SnX crystals, however does not explain formation of the stereochemically active lone pairs in GeX compounds and their absence in PbXs. An attempt of justification of the phenomenon was made by Smiles *et al.*,[20] who proposed that the atomic radius and bond lengths may affect the resulting favourable structure. The theoretical study performed by the authors for GeS and GeSe, simulating the influence of the bond length on the optimized geometry by varying the unit cell volume, did not resolve the issue, however other evidence that the explanation might be valid can be found in the literature. The XRD studies of pressure induced structural phase transition of PbX crystals revealed a transformation from a perfectly symmetric rock-salt to distorted orthorhombic structure with space group *Cmcm* (a supergroup of *Pnma*).[62–64] Furthermore, in the investigations of the thermal expansion of GeX materials, a phase transition from the orthorhombic (GeSe) or rhombohedral (GeTe) to cubic system was observed.[65,66] GeS did not undergo any phase transition up to the melting point, indicating superior stability of the distorted structure.[66] The observed behaviour of both Pb and Ge chalcogenides is in-line with the hypothesis that the interatomic distance may influence the stereochemical activity of the lone electron pairs. In the case of the former, increasing the hydrostatic pressure leads to effective reduction of the interatomic distance and induce stronger interaction between Pb and X orbitals, resulting in the crystal lattice distortion. For the latter, the thermal expansion causes an opposite effect, weakening the interaction and observed as a phase transition to a structure with higher symmetry. Considering the Sn chalcogenides, it is worth mentioning that in the high-temperature regime (above ~800 K) the *Cmcm* phase appears, although the transition is only related to the relative change of the lattice parameters (at the transition temperature $a = b$, resulting in higher symmetry of the structure) and does not involve major atomic rearrangement.[67] The result confirms the trends predicted based on the atomic orbital energies regarding the relation between the stability of the $s^2$ lone pairs and the energy distance between M $s$ and X $p$ valence states.

**Table 2.** Ionization potential, work function, electron affinity, and room temperature energy gap determined for SnS, SnSe, GeS, and GeSe. The values for GeX crystals are adapted from Ref. [59]

|      | *IP* (eV) | *φ* (eV) | *χ* (eV) | $E_g$ (eV) |
|------|-----------|----------|----------|------------|
| SnS  | 4.87      | 4.54     | 3.78     | 1.09       |
| SnSe | 4.92      | 4.80     | 4.03     | 0.89       |
| GeS  | 5.70      | 5.32     | 4.11     | 1.59       |
| GeSe | 5.47      | 5.27     | 4.27     | 1.20       |

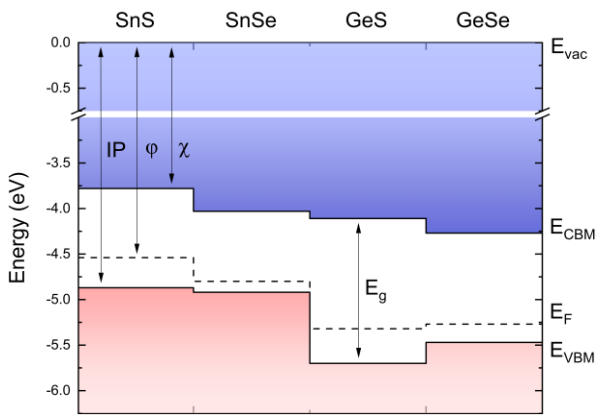

**Fig. 6.** The band alignment diagram for the group IV-IV chalcogenides SnS, SnSe, GeS, and GeSe, with respect to vacuum energy $E_{vac}$. The values for GeX crystals are adapted from Ref. [59].

## Conclusions

In conclusion, the optical properties of SnS and SnSe van der Waals crystals were investigated by complementary methods of optical spectroscopy (modulated reflectance and optical absorption) combined with *ab initio* calculations of the electronic band structure and density of states. The experimental studies confirmed the predicted indirect character of the fundamental band gap of SnS (1.09 eV at room temperature) and SnSe (0.89 eV), and revealed strong linear dichroism of the energetically higher direct optical transitions.

The anisotropy of the optical properties originates from the electronic band structure and orbital composition of the individual bands, determining the selection rules and transition probability dependent on the incident light polarization angle.

By means of UV and X-ray photoemission spectroscopy the valence bands of SnS and SnSe were examined experimentally, remaining in good agreement with the simulations based on the calculated density of states corrected by the photoionization cross-section of the individual orbitals. The measurements also allowed to determine the work functions, ionization potentials and electron affinities of the materials, providing information about the band alignment with other semiconductors for heterostructures engineering. The obtained relatively low values of the ionization potential (4.87 eV and 4.92 eV for SnS and SnSe, respectively) can be attributed to the presence of the stereochemically active Sn 5$s$ lone electron pairs in the crystal structure. Their formation is governed by the interaction between the unoccupied Sn 5$p$ orbitals and the anti-bonding state of hybridized Sn 5$s$ and S or Se valence $p$ orbitals. Along with SnX, we discuss the phenomenon and its consequences also for other group IV monochalcogenides.

In general, our results explain the origins of the linear dichroism of the optical properties of SnS and SnSe and point towards potential applications in polarization-sensitive photodetectors, but also possibilities of integration with other crystals to form van der Waals heterostructures.

## Methods

### Experimental details

The investigated samples were SnS and SnSe single crystals synthesized using modified Bridgman technique using 6N Sn, S and Se precursors. Prior to synthesis, as received precursors were purified using float zone technique to reach 6N purity. Stoichiometric ratio of Sn and chalcogen were mixed in nugget form into Bridgman ampoule with a sharp tip to limit nucleation density and enable single crystal SnS and SnSe. Typical growth temperature was set above the melting point of SnS and SnSe respectively, usually at 950 °C. The crystallization occurred by lowering rate of 1 mm per week from 950 °C hot zone to 300 °C cold zone.

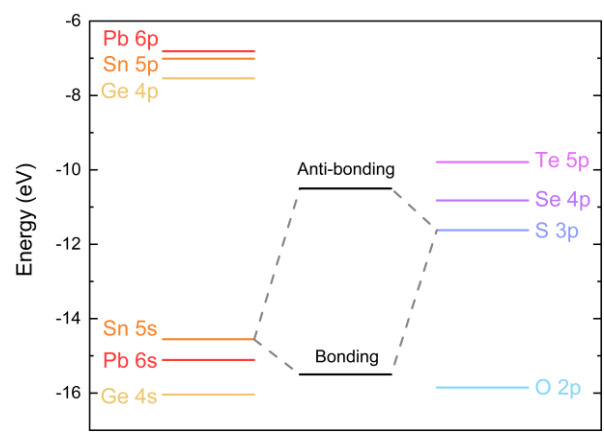

**Fig. 7.** Diagram of the electronic configuration energies of the atomic valence $s$ and $p$ orbital for group IV elements Ge, Sn, and Pb, and group VI elements O, S, Se, and Te. The orbital energies are adapted from Ref. [61]. The hybridyzed bonding and anti-bonding states are only presented for the illustation of the phenomenon and does not correspond to the values on the energy scale.

For structural characterization, X-ray diffraction (XRD) measurements were performed on a Marvel Panalytical Empyrean diffractometer in a Bragg-Brentano configuration using a Cu $K_{\alpha 1}$ $\lambda$=1.540598 Å x-ray tube and a Pixcel3D detector. A small piece of bulk crystal was placed on a Si/SiO2 substrate and a diffraction curve was recorded for it. It was measured between 10 and 75 degrees under normal conditions.

For the optical spectroscopy measurements (photoreflectance and optical absorption) the samples were exfoliated to obtained clear surface and mounted on a copper carrier, ensuring good thermal contact, inside a cryostat coupled with a closed-cycle helium cryocooler. The experiments were carried with the use of a dedicated optical setups, composed of a quartz tungsten halogen lamp (a probing white light source), a 550 mm focal length grating monochromator and a detection system, including Si or InGaAs photodiode and a lock-in amplifier (Stanford Research Systems SR830). For the photoreflectance measurement the built-in electric field modulation was achieved by periodic excitation with a 532 nm continuous wave laser, mechanically chopped at the frequency of ~300 Hz. For polarization-resolved optical absorption measurements a Glan-Taylor calcite linear polarizer and an achromatic half-wave plate were placed in the optical axis.

The photoemission studies were performed using two techniques, exploiting different radiation sources. For both variants the samples were exfoliated under ultra-high vacuum condition inside a preparation chamber. UV photoemission spectroscopy experiments were carried with the use of the synchrotron radiation (excitation energy of 100 eV) at the URANOS beamline of the SOLARIS National Synchrotron Radiation Centre (Kraków, Poland). The photoemission signal was detected with Scienta-Omicron DA30-L electron analyzer. The measurements were performed at the temperature of 77 K (achieved in a flow-type liquid nitrogen cryostat) and base pressure below $5 \times 10^{-11}$ mbar. X-ray photoemission spectroscopy (including core-level XPS) exploiting monochromatic Al $K_{\alpha}$ line (1486.6 eV) were carried at room temperature. In this technique, the photoelectrons were detected with a hemispherical analyzer Argus CU.

**Computation details**

*Ab initio* calculations were performed within the framework of density functional theory (DFT) with the use of the relativistic projector-augmented waves (PAW) datasets[68] in Vienna Ab Initio Simulation Package (VASP).[69] The Perdew-Burke-Ernzerhof (PBE) parametrization of generalized gradients approximation (GGA) to the exchange-correlation functional was employed.[70] Monkhorst-Pack k-point grid of 12 x 12 x 3, plane wave energy cutoff of 600 eV, and a semi-empirical DFT-D3 correction for vdW interactions were used.[71] The electronic band structures were calculated with the use of Heyd-Scuseria-Ernzerhof (HSE06) hybrid functional.[72]

## Conflicts of interest

The authors declare no conflicts of interests.

## Acknowledgements

This work was supported by the National Science Centre (NCN) Poland OPUS 23 no. 2022/45/B/ST7/02750. The publication was partially developed under the provision of the Polish Ministry of Science and Higher Education project "Support for research and development with the use of infrastructure of the National Synchrotron Radiation Centre SOLARIS" under contract no. 1/SOL/2021/2. The authors acknowledge the SOLARIS Centre for the access to the Beamline URANOS, where the measurements were performed. Computational studies were supported by WCSS and PLGrid Infrastructure. S.A.T. acknowledges primary support from DOE-SC0020653 (materials synthesis), Lawrence Semiconductors, NSF CMMI 2129412, NSF DMR 2330110, and NSF CMR 2111812.

# Supporting Information

# Linear dichroism of the optical properties of SnS and SnSe van der Waals crystals

*Agata K. Tołłoczko,*[*a] *Jakub Ziembicki,*[a] *Miłosz Grodzicki,*[ab] *Jarosław Serafińczuk,*[a] *Seth A. Tongay,*[c] *Natalia Olszowska,*[d] *Marcin Rosmus,*[d] *and Robert Kudrawiec*[ab]

[a] Department of Semiconductor Materials Engineering,
Wrocław University of Science and Technology,
Wybrzeże Wyspiańskiego 27, 50-370 Wrocław, Poland

[b] Łukasiewicz Research Network – PORT Polish Center for Technology Development,
Stabłowicka 147, Wrocław, Poland

[c] School for Engineering of Matter, Transport and Energy Arizona State University
Tempe, AZ 85287, USA

[d] Solaris National Synchrotron Radiation Centre, Jagiellonian University, Czerwone Maki 98, 30-392 Kraków, Poland

*Corresponding author.
E-mail address: agata.tolloczko@pwr.edu.pl

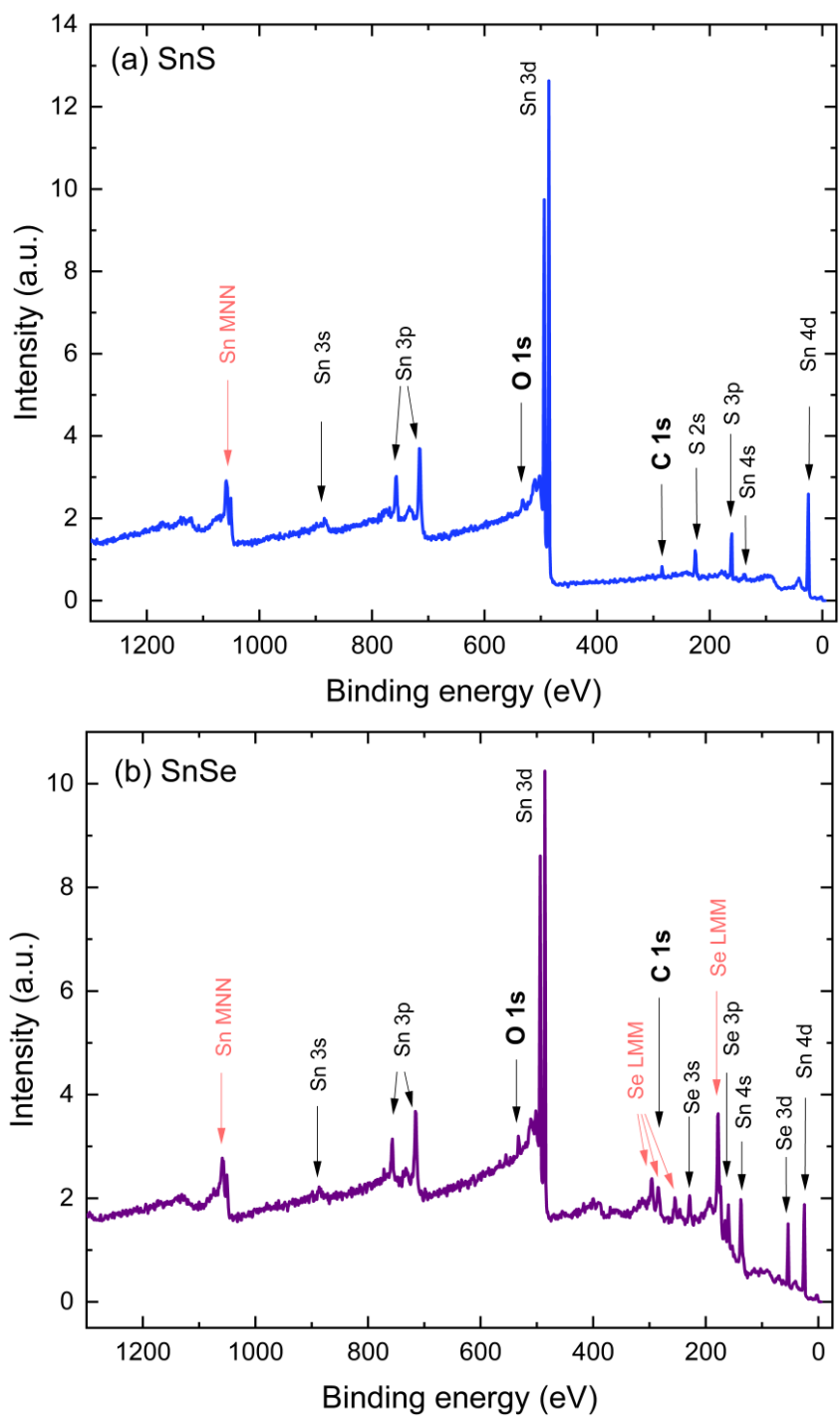

**Fig. S1.** Full-scale core-level XPS spectra acquired for SnS (a) and SnSe (b). Lines corresponding to individual atomic orbitals are identified. Auger lines are marked with red arrows and labels. Along with the signal originating from Sn and S or Se, weak contribution of O 1s (at the biding energy of 532.5 eV) and C 1s (284.7 eV) states was observed. The spectra are interpreted with regard to Ref. [S1].

**Table S1.** Binding energies of the core-level XPS lines (with SO splitting given in parentheses) measured for SnS and SnSe.

| XPS line (SO splitting) | SnS | SnSe |
|---|---|---|
| | \multicolumn{2}{c}{Binding energy (eV)} | |
| S $2p_{3/2}$ | 161.2 (1.1) | - |
| Sn $3d_{5/2}$ | 485.8 (8.5) | 485.7 (8.4) |
| Se $3d_{5/2}$ | - | 53.7 (0.9) |

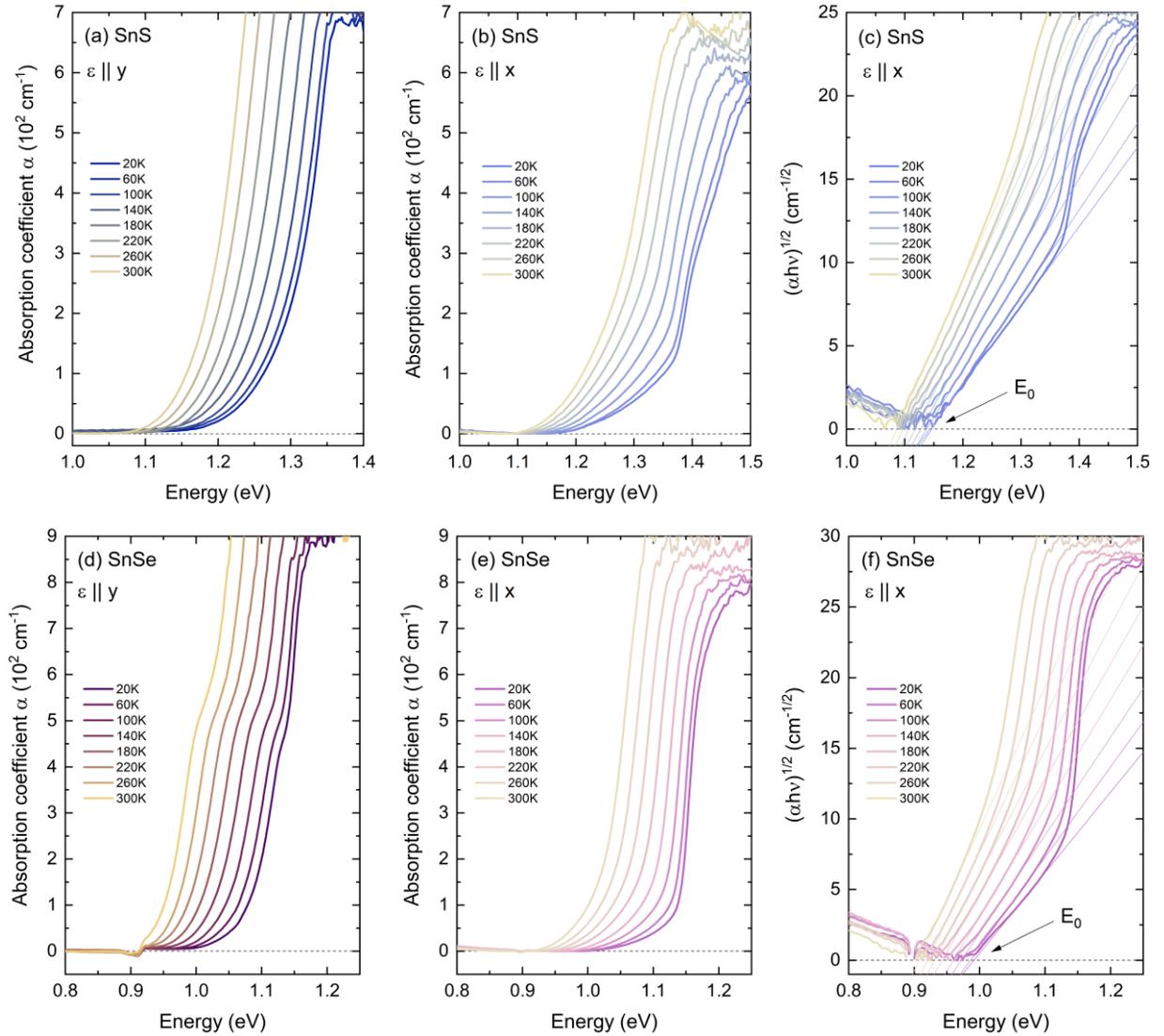

**Fig. S2.** Temperature dependence of the optical absorption measured of SnS (a,b) and SnSe (d,e) with incident light polarized along $y$ (a, d) and $x$ (b,e) in-plane crystallographic directions. Tauc plots $h\nu$ vs. $(\alpha h\nu)^{1/2}$ of the $x$-polarized absorption, allowing to determine the fundamental indirect band gap $E_0$ of SnS (c), and SnSe (f).

# Temperature dependence of the optical transition energy analysis

**Bose-Einstein formula**[S2]

$$E(T) = E(0) - \frac{2a_B}{\exp\left(\frac{\theta_B}{T}\right) - 1}$$ (Eq. S1)

$E(0)$ - energy at the temperature of 0 K,

$a_B$ - electron-phonon interaction strength,

$\theta_B$ - average phonon temperature.

**Varshni formula**[S3]

$$E(T) = E(0) - \frac{\alpha T^2}{\beta + T}$$ (Eq. S2)

$E(0)$ - energy at the temperature of 0 K,

$\alpha, \beta$ - semi-empirical Varshni coefficients.

**Table S2.** Temperature coefficients of the energies of the fundamental indirect ($E_0$) and lowest direct ($E_1$) optical transitions, determined from fitting the dependencies with Bose-Einstein (Eq. S1) and Varshni (Eq. S2) formulas.

|  | Transition | Bose-Einstein | | | Varshni | | |
|---|---|---|---|---|---|---|---|
|  |  | $E(0)$ (eV) | $a_B$ (meV) | $\theta_B$ (K) | $E(0)$ (eV) | $\alpha$ ($10^{-4}$ eV K$^{-1}$) | $\beta$ (K) |
| SnS | $E_0$ | 1.15 | 19.2 | 140 | 1.15 | 3.02 | 110 |
|  | $E_1$ | 1.23 | 27.7 | 133 | 1.23 | 4.67 | 112 |
| SnSe | $E_0$ | 0.99 | 28.4 | 133 | 0.99 | 4.85 | 119 |
|  | $E_1$ | 1.06 | 78.6 | 223 | 1.07 | 8.75 | 230 |

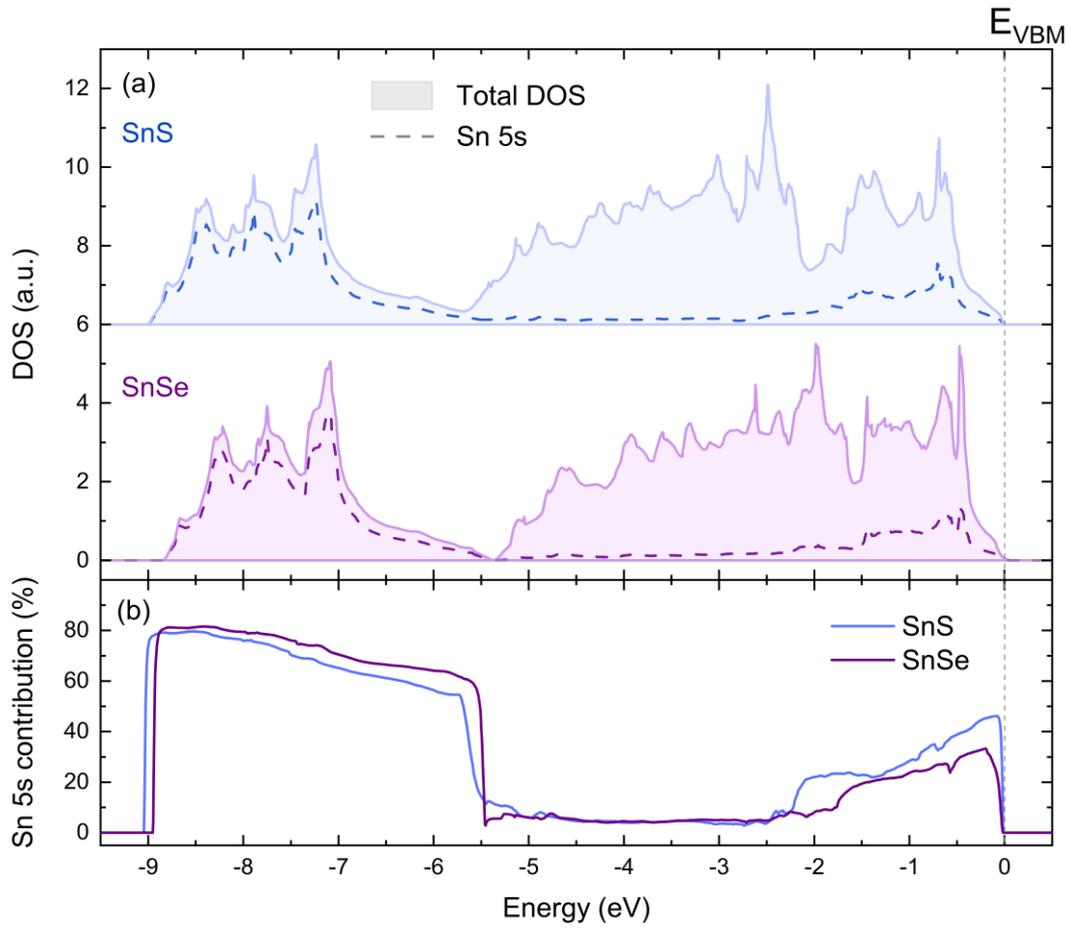

**Fig. S3.** (a) Total (shaded areas) and partial Sn 5s (dashed lines) DOS calculated for SnS (top plot) and SnSe (bottom plot). (b) The percentage contribution of the Sn 5s orbital to total DOS.

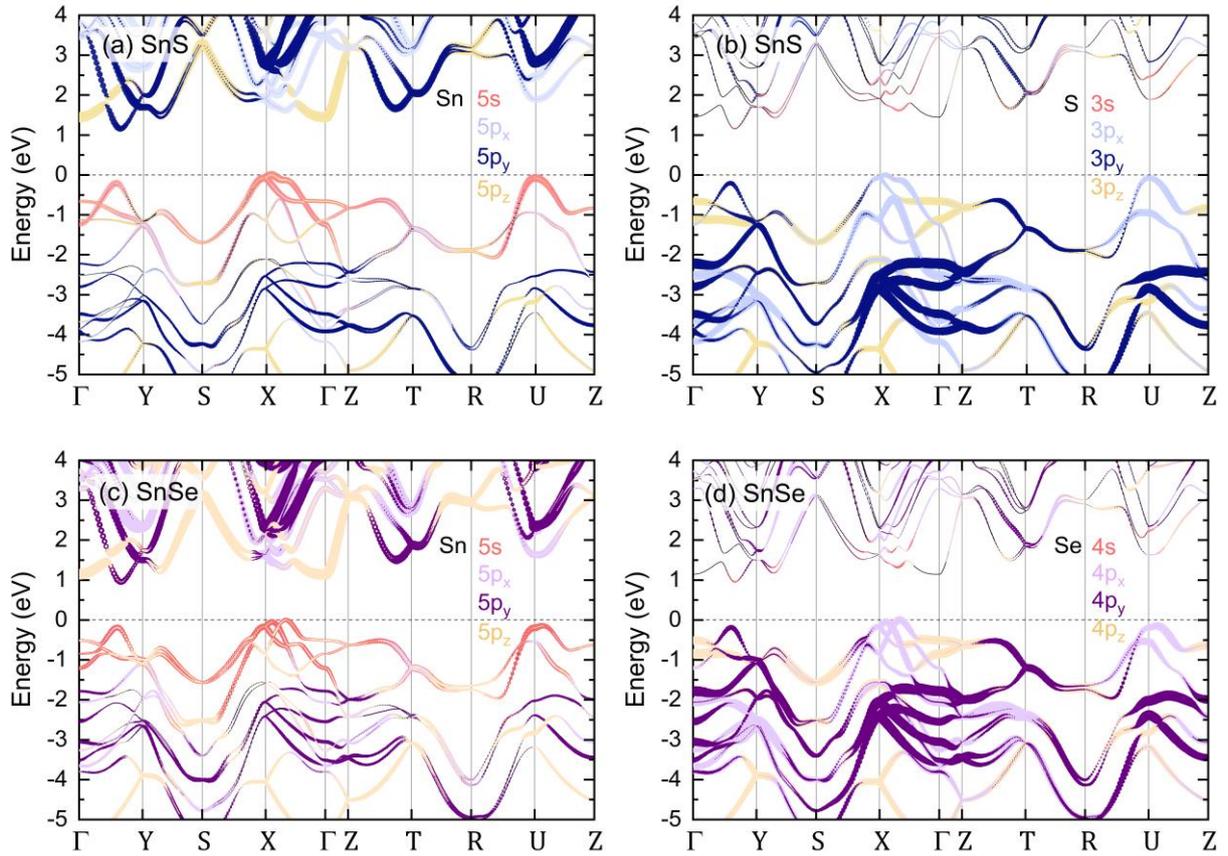

**Fig. S4.** The electronic band structure calculated with the use of mBJ exchange potential for SnS (a,b) and SnSe (c,d), with superimposed contribution of the valence *s* and *p* (with three spatial components $p_x$, $p_y$, $p_z$) orbitals of Sn (a,c), S (b) and Se (c) atoms, illustrated with the size of the plot points.